\documentclass[twocolumn,showpacs,preprintnumbers,amsmath,amssymb,prl]{revtex4}
\usepackage[dvips]{graphicx}
\usepackage{dcolumn}
\usepackage{bm}

\begin{document}

\preprint{APS/123-QED}

\title{Photo-induced precession of magnetization in ferromagnetic (Ga,Mn)As}

\author{Y. Hashimoto}
\email{hashimoto.y.ad@m.titech.ac.jp}
\affiliation{Imaging Science and Engineering Laboratory, Tokyo Institute of Technology, 4259-G2-13 Nagatsuta, Midori-ku, Yokohama 226-8502, Japan}
\author{H. Munekata}
\affiliation{Imaging Science and Engineering Laboratory, Tokyo Institute of Technology, 4259-G2-13 Nagatsuta, Midori-ku, Yokohama 226-8502, Japan}
\date{\today}

\begin{abstract}
Precession of magnetization induced by pulsed optical excitation is observed in a ferromagnetic semiconductor (Ga,Mn)As by time-resolved magneto-optical measurements. It appears as complicated oscillations of polarization plane of linearly-polarized probe pulses, but is reproduced by gyromagnetic theory incorporating an impulsive change in an effective magnetic field due to change in magnetic anisotropy. The shape of the impulse suggests significant \textit{non-thermal} contribution of photo-generated carriers to the change in anisotropy through spin-orbit interaction.
\end{abstract}
\pacs{75.50.Pp, 76.50.+g, 78.20.Ls, 78.47.+p}
\maketitle

   Carrier-mediated ferromagnetism in III-V-based ferromagnetic semiconductors has provided an opportunity for fundamental studies of manipulating magnetization $M$ through carrier- and orbital-mediated mechanisms~\cite{Die01, Pap07}.  Coherent control of $M$ via the excitation of an electronic system is especially of great interest for ultrafast spin-electronics/photonics and quantum computing.  Light-induced precession of $M$ of \textit{non-thermal} origin~\cite{Cho06}, however, has not yet been uncovered in this class of semiconductors, except the one caused by the light-induced lattice/spin heating~\cite{DMWang07}.  Not only for semiconductors but also for magnetic materials, laser-induced precession and switching of spins~\cite{Kamp02, Han05, Stan07} has been of great importance for further developing magnetic storage technology.  Among the works with magnetic materials, precession in a ferrimagnetic garnet film~\cite{Han05} was the clear example that was caused by a non-thermal mechanism: direct intra-ionic excitation with intense laser pulses followed by charge transfer among magnetic ions.  In this case, however, long lifetime of excited states would impede manipulation of spins in the time scale of current interest ($<$ 1 ns).

   This letter concerns the precession of $M$ induced by an impulsive change in magnetic anisotropy through \textit{inter-band} excitation with ultra-short optical pulses (150 fs) of a few $\mu $J/cm$^{2}$.  The shape of the impulse extracted by the model calculation based on the Landau-Lifshitz-Gilbert (LLG) equation strongly suggests that the change in anisotropy is not of sample heating~\cite{Koj03, JWang05L, Ast05, DMWang07} but is attributed to the generation, cooling, and subsequent annihilation of photo-generated holes.  Our findings suggest the presence of \textit{non-thermal} pathway of coupling spins and light, at least in semiconductor-type solids.  

\begin{figure}
\includegraphics[scale = 0.45] {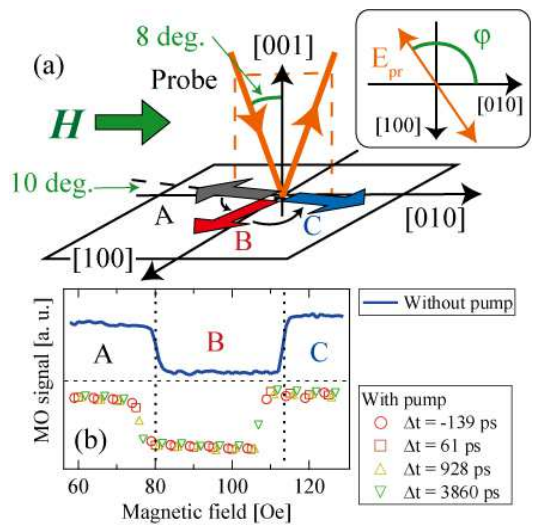}
\caption{\label{fig:Base} (color online).  (a) Schematic illustration of experimental configurations including static magnetization orientations \textit{A}, \textit{B}, and \textit{C}. $\varphi $ represents angle of $E_{pr}$ with respect to [010] axis.  (b) Magneto-optical (MO) signals observed while sweeping magnetic field.  Upper and lower plots are those obtained without and with pump pulses, respectively.} 
\end{figure}

   A 100-nm thick Ga$_{1-x}$Mn$_{x}$As sample used in this study was grown by molecular beam epitaxy on a GaAs [001] substrate at the substrate temperature of 235 $^{\circ }$C. Curie temperature, Mn content, and hole-concentration are $T_{c}$ $\sim $ 45 K, $x \sim $ 0.02, and $p \sim 1.8 \times 10^{20}cm^{-3}$, respectively. The magnetization easy axis lies in the plane, as verified by separate magnetization measurements~\cite{Tak06}. The sample was set in a cryostat equipped with an electromagnet and kept at 10 K. A single-wavelength pump and probe technique was used for investigating the optically-induced spin dynamics through magneto-optical effects.  The light source was a Ti:Sapphire laser with a pulse width of 150 fs and a repetition rate of 76 MHz.  The experimental configuration is shown schematically in Fig.~\ref{fig:Base}(a).  Both pump and probe beams were linearly polarized, and were focused on the sample surface through a single lens into a spot diameter of about 150 $\mu $m.  The pump fluence was around 3.4 $\mu $J/cm$^{2}$ per pulse, with the ratio of pump and probe intensities 10:1.  Rotation angle of the probe beam polarization was detected by means of an optical bridge circuit with the resolution of better than 10 $\mu $deg.  The photon energy of the laser was set at 1.569 eV (790 nm) unless noted differently. 

   When magneto-optical (MO) signals were measured without pump pulses under an external magnetic field along the [010] axis [Fig.~\ref{fig:Base}(a)], a double-step feature is obtained, as shown in Fig.~\ref{fig:Base}(b). This is typical of the two-step magnetization rotation process observed via magnetic linear dichroism (MLD)~\cite{Moo03}.  Steps \textit{A}, \textit{B}, and \textit{C} in the profiles shown in Fig.~\ref{fig:Base}(b) correspond to magnetization orientations nearly along [0$\bar {1}0$], [100], and [010], respectively, as denoted in Fig.~\ref{fig:Base}(a).  Rigorously stated~\cite{Moo03}, the orientation of $M$ in the state \textit{A} is in the direction that is 10 degree off the [010] axis toward the [110] axis, reflecting the contribution of the [110] uniaxial anisotropy~\cite{Saw05}.  When pump pulses were added, the entire double-step profile shifted toward lower magnetic fields by about 3 Oe [Fig.~\ref{fig:Base}(b)] for the time delay between $t$ = +60 ps and -140 ps. The same amount of shift was also detected at $t$ as short as +3 ps. These facts indicate that the dissipation rate of thermal energy from an excited spot was slower than the interval of a pulse train (13 ns), and heat was accumulated in the spot. On the basis of the observed shift of 3 Oe, the temperature increase is estimated to be 0.9 K, which includes slight heating and subsequent cooling within the detection limit of 0.1 K in between the pulse excitation. 

\begin{figure}
\includegraphics[scale = 0.65] {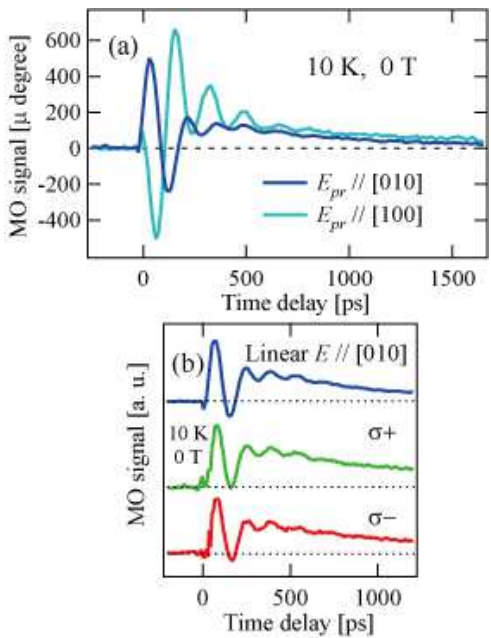}
\caption{\label{fig:TRMOSig} (color online).  (a) Temporal profiles of MO signals obtained at 10 K with initial magnetization orientation \textit{A} [Fig.~\ref{fig:Base}(a)] for two different polarizations.  The polarizations of pump and probe pulses were set parallel.  No external magnetic field is applied.  (b) MO profiles obtained for three different polarizations of the pump pulse with a fixed probe beam polarization of $E_{pr}$ // [010].}
\end{figure}

\begin{figure}
\includegraphics[scale = 0.45] {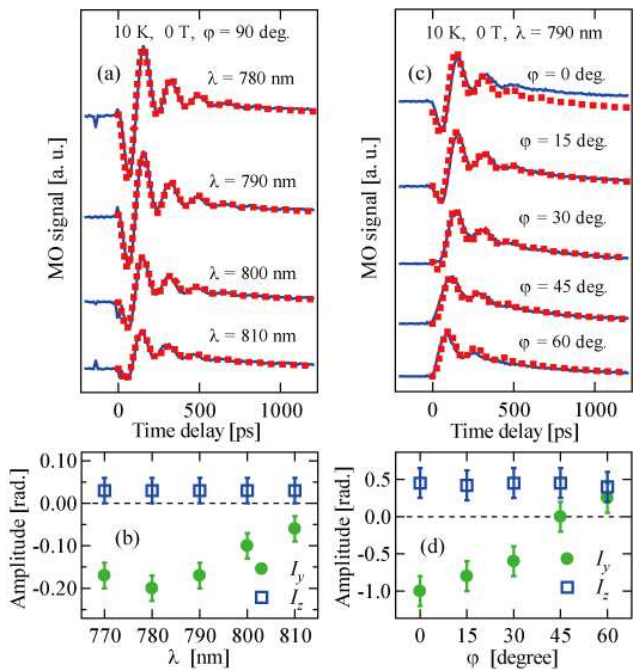}
\caption{\label{fig:WLAngle} (color online).  (a) MO profiles (blue lines) obtained with initial magnetization orientation \textit{A} [Fig.~\ref{fig:Base}(a)] for four different wavelengths $\lambda $, together with calculated profiles (red dots).  Polarization plane of a probe beam is set at $\varphi $ = 90$^{\circ }$.  (b) Plots of $I_{y}$ and $I_{z}$ used for calculated profiles shown in (a) as a function of $\lambda $.  (c) MO profiles (blue lines) obtained with initial magnetization orientation \textit{B} [Fig.~\ref{fig:Base}(a)] for five different $E_{pr}$ ranging from $\varphi $ = 0$^{\circ }$ to 60$^{\circ }$, together with calculated profiles (red dots).  Polarization planes of a pump and probe beams are set parallel.  (d) Plots of $I_{y}$ and $I_{z}$ used for calculated profiles shown in (c) as a function of $\varphi $ of $E_{pr}$.}
\end{figure}

   Profiles of time-resolved magneto-optical signals obtained at the initial magnetization orientation \textit{A} [Fig.~\ref{fig:Base}(a)] with two different polarizations of probe pulse, namely $E_{pr}$ // [100] ($\varphi $ = 90$^{\circ }$, a light blue line) and $E_{pr}$ // [010] ($\varphi $ = 0$^\circ $, a blue line), are shown in Fig.~\ref{fig:TRMOSig}(a).  Both profiles exhibit oscillations and exponential decays.  The oscillations are primarily of the bulk-like mode according to the Ref.~\cite{DMWang07}.  Between the two profiles, a phase shift and a change in amplitude are noticeable. These complexities are beyond a simple MLD picture, which we discuss later in the context of detailed experimental data shown in Figs.~\ref{fig:WLAngle}(a) and (c).  A linear increase in oscillation frequency with an external magnetic field and disappearance of the oscillation at around the Curie temperature~\cite{Tak06} both indicate that the observed MO signals are due to precession of magnetization.  The origin of oscillation has also been confirmed by separate ferromagnetic resonance experiment~\cite{Mat06}.

A separate study on dynamics of photo-carriers in (Ga,Mn)As using a two-color pump-and-probe techniques has suggested that, because electrons are trapped quickly in deep centers (10 ps), the lifetime of photo-holes is 200 ps or longer~\cite{Mitsu}. The exponential decays shown in Figs.~\ref{fig:TRMOSig}(a) and (b) are not due to artifacts associated with a change in the sample temperature but are most likely originated from the photo-holes.

   It is worth noting that the profile is almost independent of the polarization of the pump pulse, as shown in Fig.~\ref{fig:TRMOSig}(b).  This fact reveals that the oscillation is not associated with the angular momentum of circularly polarized excitation. 

   A series of temporal profiles taken with configuration \textit{A} using various pump-and-probe wavelengths $\lambda $ is shown in Fig.~\ref{fig:WLAngle}(a).  The polarization of the probe pulses is fixed at $\varphi $ = 90$^{\circ }$.  The oscillation amplitude decreases with increasing $\lambda $, whereas phase and frequency remain nearly unchanged.  Another series measured with configuration \textit{B} for various $\varphi $ is shown in Fig.~\ref{fig:WLAngle}(c).  Note that $\varphi $ = 0$^\circ $ represents $E_{pr} \perp $ $M$ in this case. A reduction of amplitude is again noticeable with increasing $\varphi $.  Taking into account that the contribution of longitudinal Kerr rotation (LKR)~\cite{Hra02} is negligibly small when $E_{pr} \perp $ $M$, we are able to interpret the observed signals as being due to both MLD~\cite{ Moo03, Kim05L} (the in-plane motion of $M$) and polar Kerr rotation (PKR)~\cite{Hra02, Lang05} (the out-of-plane motion of $M$).

   The shape of a magnetic impulse induced by optical excitation was studied by the model calculation based on classical gyromagnetic theory.  The LLG equation was solved numerically~\cite{RK} with various trial functions $H(x, y, z, t)$ and damping constant $\alpha $, and the calculated $M$(t) $\equiv $ ($M_{x}$(t), $M_{y}$(t), $M_{z}$(t)) was put into the phenomenological equation 
\begin{equation}
\label{eq:fit}
\Theta _{fit}(t) = I_{x} \frac{M_{x}(t) - M_{s}}{M_{s}} + I_{y} \frac{M_{y}(t)}{M_{s}} + I_{z} \frac{M_{z}(t)}{M_{s}}
\end{equation}
Here, $I_{x}$ and $I_{y}$ are phenomenological magneto-optical (MO) constants associated with MLD, and $I_{z}$ is the constant associated with PKR.  $M_{s}$ is the magnetization of the sample.  $x$, $y$, and $z$ axes are set in such a way that $x$ // $M(initial)$, $y \perp  x$ in the sample plane, and $z \perp $ $x$-$y$ plane.  The use of the LLG equation instead of the Bloch equation is justified by the low excitation power which produced 10$^{15}$ -10$^{16}$ cm$^{-3}$ photo-generated carriers per pulse.  This value is four to five orders of magnitude smaller than the background carrier concentration of 1.8 $\times$ 10$^{20}$cm$^{-3}$, which causes little impact on $|M|$.  Note that the direction of the tilt can not be determined explicitly by Eq.~\ref{eq:fit}, unless relative magnitudes of $I_{n}$ are determined independently. This problem was circumvented by carefully checking wavelength and probe-polarization dependencies of $I_{n}$, as described later.

That the amplitude of the first oscillation is merely larger than subsequent oscillations suggests that duration of a light-induced magnetic impulse is longer than one-half of a precession period.  Simulation has been carried out with this picture in mind, and, as shown by the red dots in Figs.~\ref{fig:WLAngle}(a) and (c), we found that a magnetic impulse having a modest rise and decay, represented by the combination of Eqs.~\ref{eq:Heffa} and \ref{eq:Heffb}, reproduced experimental data.  Eq.~ \ref{eq:Heffa} represents the shape of the temporal profile, whereas Eq.~ \ref{eq:Heffb} specifies the direction of tilt.  For simplicity, we only considered a tilt occurring toward either $y$ or $z$ direction.
\begin{subequations}
\label{eq:Heff}
\begin{equation}
\label{eq:Heffa}
\theta (t) = \theta _{0} (1 - exp(- t/ 50 ps)) exp(- t/ 500 ps) 
\end{equation}
\begin{equation}
\label{eq:Heffb}
H_{eff}(t)=(H_{0}cos\theta (t),0,H_{0}sin\theta (t))\simeq(H_{0},0,H_{0}\theta (t))
\end{equation}
\end{subequations}
  $H_{0}$ is the effective magnetic field in the initial magnetic states \textit{A} and \textit{B}, $\theta$ ($\ll \pi$) the temporal change in the angle of the effective magnetic field with respect to the $x$ axis, and $\theta _{0}$ = 10 mrad the adjusting parameter scaled by the product of an actual tilt angle and $I_{n}$.  $\mu _{0}H_{0}$ = 0.21 $\pm $ 0.01 Tesla and the damping parameter $\alpha$ = 0.16 $\pm $ 0.02 were used, which were consistent with those for non-annealed samples \cite{Welp03, Liu03, Sin04}. 

   It is known that MO constant of MLD ($I_{x}$ and $I_{y}$) exhibits strong dependencies on both $\lambda $ and $E_{pr}$ of an optical pulse~\cite{Kim05L}, whereas that of PKR ($I_{z}$) is almost independent of $E_{pr}$.  With these criteria, magneto-optical constants extracted from each fit could be labeled with $I_{x}$, $I_{y}$, and $I_{z}$.  The results are shown in Figs.~\ref{fig:WLAngle}(b) and (d), in which the difference in $\lambda$ and $E_{pr}$ dependencies is clearly seen between $I_{y}$ and $I_{z}$~\cite{EX}.  The contribution from the $x$ component is found negligibly small, reflecting a non-linear $\varphi $ dependence of MLD.  Plots in Fig.~\ref{fig:WLAngle}(b) follow the relation $\delta A_{PKR}/\delta (h\nu ) \propto A_{MLD}$~\cite{Kim05L} in which $A_{MLD}$ and $A_{PKR}$ are the magnitudes of MLD and PKR.  $I_{y}$ = 0 at $\varphi $ = 45$^{\circ }$ seen in Fig.~\ref{fig:WLAngle}(d) is also a distinct character of MLD.  Success in labeling MO constants allows us to determine backward that the $z$ component in Eq.~\ref{eq:Heffb} is altered by the pulsed excitation.  It should be noted, however, that the tilt of $H_{eff}$ towards $z$ direction, which by itself is out of our expectation, relies on how $I_{n}$ was determined, and thus tentative.  A model allowing a tilt toward the $y$-$z$ direction should also be tested to precisely nail down the direction of a tilt.

\begin{figure}
\includegraphics[scale = 0.6] {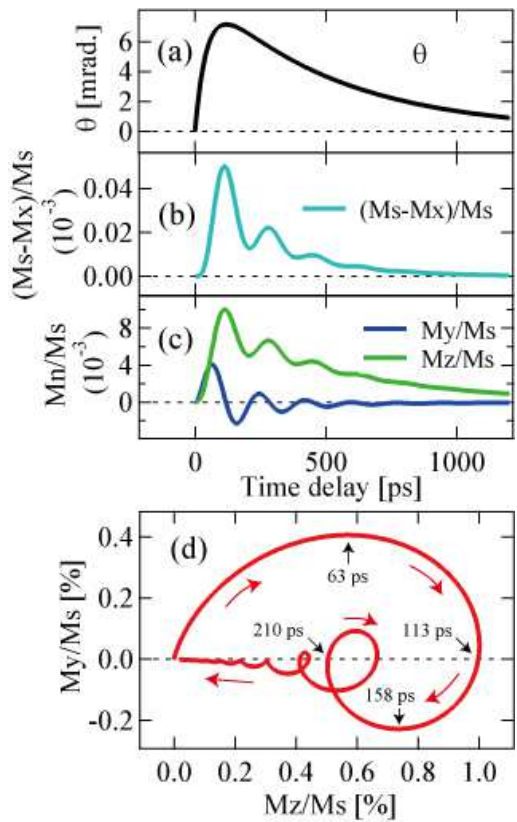}
\caption{\label{fig:Fitting} (color online).  Temporal profiles of, (a) the angle of tilt $\theta$ of $H_{eff}$, (b) $x$ component of magnetization, and (c) $y$ and $z$ components of magnetization $M_{y}$ and $M_{z}$.  These profiles are extracted on the basis of experimental curves shown in Figs.~\ref{fig:TRMOSig} and~\ref{fig:WLAngle}.  (d) Precessional motion of magnetization in $M_{y}$ - $M_{z}$ parameter space.}
\end{figure}

   The shape of magnetic impulse is shown in Fig.~\ref{fig:Fitting}(a), together with ($M_{x}$(t), $M_{y}$(t), $M_{z}$(t)) in Figs.~\ref{fig:Fitting}(b) and (c).  Among the three oscillatory components, the amplitude of $M_{x}$ is two orders of magnitude smaller than $M_{y}$ and $M_{z}$, reflecting a small tilt angle of $H_{eff}$.  As a whole, optically induced oscillatory MO signals can be viewed as precession of $M$ in the $y$-$z$ plane [Fig.~\ref{fig:Fitting}(d)].  Upon pulsed excitation, $H_{eff}$ is tilted with $M$ following the tilted $H_{eff}$ for the first 120 ps.  The tilted $H_{eff}$ thereafter turns into relaxation, whereupon free precession takes place with damping until $M$ returns to its initial magnetic state. A long tail in MO signals is explained as reminiscence of the $M_{z}$ component.

  If an impulsive change in magnetic anisotropy came from a slight increase in sample temperature, it would also manifest itself as a temporal change in $H_{eff}$.  However, the observed free precession is reproduced by the single $ \mu _{0} H_{0}$ value within the accuracy of 0.01 Tesla.  Moreover, the fact that oscillation frequency did not change with pump power ($\leq $ 3.4 $\mu J/cm^{2}$, data not shown) strongly suggests that thermal influence of pulsed excitation is not responsible.  We therefore come to the conclusion that the observed precession is induced by the \textit{non-thermal} influence of pulsed excitation.

  An increase in tilt angle that takes place in the first 120 ps suggests the presence of a \textit{post-photocarrier process}.  Hot photo-carriers (photo-holes) generated by the optical excitation are cooled down via emission/absorption of acoustic phonons in about 100 ps in the case of nearly resonant excitation~\cite{Shah99}.  During this process, the number of thermally cooled holes near the equilibrium Fermi level increases, which in tern alters magnetic anisotropy via spin-orbit interaction, in the time scale of energy relaxation of carriers.  Referring to the equilibrium picture~\cite{Die01, Saw05}, a change in the numbers of holes alters the balance between cubic and in-plane uniaxial anisotropies, which would result in the tilt of $H_{eff}$ presumably toward $z$ axis.  Extension of \textit{p-d} Zener model~\cite{Die01} is desired.  The decay profile of $\theta$ can be understood in terms of annihilation of cooled photo-holes from the valence band.  The shape of a decay curve could be manipulated by controlling defects in the samples.  

   In summary, photo-induced precession of magnetization in ferromagnetic (Ga,Mn)As has been studied.  Complicated oscillatory MO profiles were reproduced successfully by gyromagnetic theory connected with magnetic linear dichroism (MLD) and polar Kerr rotation (PKR) which respectively detects in-plane and out-of-plane magnetization components.  It is found that the pulsed excitation yields an impulsive change in magnetic anisotropy for which \textit{non-thermal} influence of optical excitation is responsible.  Our findings suggest that hole cooling is presently one of the limiting factors for ultra-fast control of carrier-/orbital-mediated ferromagnetism through pulsed excitations with light as well as, presumably, an electric field.

   The authors are grateful to Dr. A. Oiwa of the University of Tokyo for his collaboration in sample preparation and measurements. This work is supported in part by Grant-in-Aid for Scientific Research from JSPS (No. 17206002), MEXT (No.19048020), and by the NSF-IT program (DMR-0325474).


\begin{thebibliography}{24}
\bibitem[{RK()}]{RK}
\bibinfo{note}{The Runge-Kutta method was used for the calculations.}
\end{thebibliography}
\end{document}